\documentclass[twocolumn,prl,showpacs,preprintnumbers,superscriptaddress]{revtex4}
\usepackage[english]{babel}

\usepackage{amsmath}
\usepackage{dcolumn}
\usepackage{bm}
\usepackage{epsfig}
\usepackage{epstopdf}
\usepackage{graphicx}
\usepackage{amsfonts}
\usepackage{amssymb}
\usepackage{mathrsfs}
\usepackage{color}
\usepackage{appendix}
\usepackage{multirow}
\usepackage{graphicx}
\usepackage{longtable}
\usepackage{afterpage}
\usepackage{rotating}
\begin{document}

\title{Measurement of infrared magic wavelength for an all-optical trapping of $^{40}$Ca$^{+}$ ion clock}
\author{Yao Huang}
\affiliation{State Key Laboratory of Magnetic Resonance and Atomic and Molecular Physics, Wuhan Institute of Physics and Mathematics, Chinese Academy of Sciences, Wuhan 430071, China}
\affiliation{Key Laboratory of Atomic Frequency Standards, Wuhan Institute of Physics and Mathematics, Chinese Academy of Sciences, Wuhan 430071, China}
\author{Hua Guan}
\affiliation{State Key Laboratory of Magnetic Resonance and Atomic and Molecular Physics, Wuhan Institute of Physics and Mathematics, Chinese Academy of Sciences, Wuhan 430071, China}
\affiliation{Key Laboratory of Atomic Frequency Standards, Wuhan Institute of Physics and Mathematics, Chinese Academy of Sciences, Wuhan 430071, China}
\author{Chengbin Li}
\affiliation{State Key Laboratory of Magnetic Resonance and Atomic and Molecular Physics, Wuhan Institute of Physics and Mathematics, Chinese Academy of Sciences, Wuhan 430071, China}
\affiliation{Key Laboratory of Atomic Frequency Standards, Wuhan Institute of Physics and Mathematics, Chinese Academy of Sciences, Wuhan 430071, China}
\author{Huaqing Zhang}
\affiliation{State Key Laboratory of Magnetic Resonance and Atomic and Molecular Physics, Wuhan Institute of Physics and Mathematics, Chinese Academy of Sciences, Wuhan 430071, China}
\affiliation{Key Laboratory of Atomic Frequency Standards, Wuhan Institute of Physics and Mathematics, Chinese Academy of Sciences, Wuhan 430071, China}
\affiliation{University of Chinese Academy of Sciences, Beijing 100049, China}
\author{Baolin Zhang}
\affiliation{State Key Laboratory of Magnetic Resonance and Atomic and Molecular Physics, Wuhan Institute of Physics and Mathematics, Chinese Academy of Sciences, Wuhan 430071, China}
\affiliation{Key Laboratory of Atomic Frequency Standards, Wuhan Institute of Physics and Mathematics, Chinese Academy of Sciences, Wuhan 430071, China}
\affiliation{University of Chinese Academy of Sciences, Beijing 100049, China}
\author{Miao Wang}
\affiliation{State Key Laboratory of Magnetic Resonance and Atomic and Molecular Physics, Wuhan Institute of Physics and Mathematics, Chinese Academy of Sciences, Wuhan 430071, China}
\affiliation{Key Laboratory of Atomic Frequency Standards, Wuhan Institute of Physics and Mathematics, Chinese Academy of Sciences, Wuhan 430071, China}
\affiliation{University of Chinese Academy of Sciences, Beijing 100049, China}
\author{Liyan Tang}
\affiliation{State Key Laboratory of Magnetic Resonance and Atomic and Molecular Physics, Wuhan Institute of Physics and Mathematics, Chinese Academy of Sciences, Wuhan 430071, China}
\affiliation{Key Laboratory of Atomic Frequency Standards, Wuhan Institute of Physics and Mathematics, Chinese Academy of Sciences, Wuhan 430071, China}
\author{Tingyun Shi}
\affiliation{State Key Laboratory of Magnetic Resonance and Atomic and Molecular Physics, Wuhan Institute of Physics and Mathematics, Chinese Academy of Sciences, Wuhan 430071, China}
\affiliation{Key Laboratory of Atomic Frequency Standards, Wuhan Institute of Physics and Mathematics, Chinese Academy of Sciences, Wuhan 430071, China}
\author{K. Gao}
\email{klgao@wipm.ac.cn}
\affiliation{State Key Laboratory of Magnetic Resonance and Atomic and Molecular Physics, Wuhan Institute of Physics and Mathematics, Chinese Academy of Sciences, Wuhan 430071, China}
\affiliation{Key Laboratory of Atomic Frequency Standards, Wuhan Institute of Physics and Mathematics, Chinese Academy of Sciences, Wuhan 430071, China}
\affiliation{Center for Cold Atom Physics, Chinese Academy of Sciences, Wuhan 430071, China}

\date{\today}

\begin{abstract}
For the first time, we experimentally determine the infrared magic wavelength for the $^{40}$Ca$^{+}$ $4s\, ^{2}\!S_{1/2} \rightarrow
3d\,^{2}\!D_{5/2}$ electric quadrupole transition by observation of the light shift canceling in $^{40}$Ca$^{+}$ optical clock. A "magic" magnetic field direction is chosen to make the magic wavelength insensitive to both the linear polarization purity and the polarization direction of the laser. The determined magic wavelength for this transition is 1056.37(9)~nm, which is not only in good agreement with theoretical predictions but also more precise  
by a factor of about 300. Using this measured magic wavelength we also derive the differential static polarizability to be $-44.32(32)$~a.u., which will be an important input for the evaluation of the blackbody radiation shift at room temperatures. Our work paves a way for all-optical-trapping of $^{40}$Ca$^{+}$ optical clock.
\end{abstract}

\pacs{32.10.Dk, 06.20.F, 06.30.Ft, 37.10.Ty} \maketitle

With rapid development of laser technology, state-of-the-art optical clocks have now reached an accuracy or frequency stability at the level of 10$^{-18}$ or higher~\cite{Ushijima15,Huntemann16,McGrew18,Brewer19,Bothwell19}, which is two orders of magnitude better than the state-of-the-art microwave atomic clocks. At this level of accuracy, optical clocks can play a critical role in redefining the second~\cite{Targat13}, in searching for variation of fundamental constants~\cite{Huntemann14,Safronova18}, and in chronometric leveling~\cite{Grotti18}.
For many neutral-atom optical lattice clocks, the ac-Stark shift due to black body radiation (BBR) or lattice lasers~\cite{McGrew18,Bothwell19} can be a limiting factor for achieving such high accuracy~\cite{McGrew18,Bothwell19}; for ion-based clocks, on the other hand,
micromotion shifts~\cite{Huntemann16,Brewer19} may limit the accuracy of some clocks.
One way to reduce the micromotion shifts is to apply the all-optical trapping technique~\cite{Schneider10,Huber14,Lambrecht17}, where the micromotion shift will be gone when the rf field is switched off. Since the laser used for all-optical trapping can be chosen at a magic wavelength~\cite{LeBlanc07,Arora11,Herold12, Holmgren12}, the energy shift in the relevant transition will be zero and thus the trapping potential will introduce no shift in the clock transition frequency. Therefore, for a magic-wavelength optical-trapped ion, both the micromotion and ac-Stark shift can be greatly suppressed. In addition to the accuracy of a clock, the frequency stability is also a very important issue when evaluating a clock. Comparing to neutral-atom lattice clocks, the stability of a single ion clock is limited by the signal to noise ratio. Recently, the optical trapping of Coulomb ion crystals has been demonstrated~\cite{Schmidt18}, which sheds a light on the development of all-optical trapping ion clocks using multiple ions to achieve a better frequency stability

Precision measurements of magic wavelengths in atoms are also very important in fundamental studies of atomic structure. For example, a measurement of line strength ratio by magic wavelength can bring a new perspective for determining accurate transition matrix elements, which are important in testing atomic computational methods and in interpreting atomic parity non-conservation~\cite{Derevianko00,Sahoo06,Porsev09}. Precision measurements of magic wavelengths in ions can be used to derive relevant oscillator strengths and polarizabilities for clock states~\cite{Liu15}, which is essential for evaluating the BBR shift at the 10$^{-18}$ level at room temperatures.

The magic wavelengths of Ca$^+$ have recently been studied both theoretically~\cite{Tang13,Kaur15,Jiang17} and experimentally~\cite{Liu15}. Two magic wavelengths for the $4s_{1/2} \rightarrow 3d_{5/2}$ ($m_J$ = 1/2, 3/2) clock transitions near 395.79~nm have been measured to high accuracy, which  are in well agreement with all existing theoretical predictions. However, these magic wavelengths are very close to the $4s_{1/2}\rightarrow4p_{3/2}$ and $4s_{1/2}\rightarrow4p_{1/2}$ resonant transitions. The near resonant light has high spontaneous photon scattering rates that can result in a high heating process~\cite{Haycock97}. Thus, these magic wavelengths are not ideal choices for the optical trapping of the ions.
Therefore, in order to do optical trapping of ions, it is important to search for magic wavelengths far off any resonant transitions; for $^{40}$Ca$^+$ in particular, magic wavelengths in the infrared region are desirable.

In this Letter, we will report the experimental measurement of an infrared magic wavelength by observation of the light shift canceling in $^{40}$Ca$^+$ optical clock. The clock has an uncertainty of 2.2 $\times$ 10$^{-17}$ and a 10$^{-16}$ level stability at a few seconds~\cite{Huang19}. The clock is suitable for making a differential measurement, the clock uncertainty would only introduce a negligible measurement uncertainty of $<$ 0.001 nm. We will present a method to extract a reduced transition matrix element using our measured magic wavelength. We will also determine a static differential polarizability that is an important parameter in evaluating the BBR shift at room temperatures.

Calculating or measuring an infrared magic wavelength is very different from measuring a near-resonance magic wavelength~\cite{Liu15}. Briefly speaking,in theoretical calculation, the predicted magic wavelengths have much larger uncertainty compared to the  near-resonance magic wavelengths; in the experiments, for a near-resonance magic wavelength, it is much less sensitive to magnetic field direction, laser propagation direction, and laser polarization direction. For measuring a far-off-resonance magic wavelength, however, one needs to carefully control the laser and magnetic field conditions and carefully evaluate systematic shifts. To setup the experiment, first of all, a single $^{40}$Ca$^+$ ion is trapped in a miniature ring Paul trap and the temperature of ion is laser cooled to a few mK. To measure the magic wavelength, the clock laser is locked to the Zeeman components of clock transition and the light shift on the clock transition can be observed by switching on and off the laser with wavelength around 1050~nm (named L$_m$ laser for short in the following sections). To keep the L$_m$ laser linearly polarized during the measurement, a polarizer (Glan-Tyler Prism) is placed in the light path before ion-light interaction takes place. In doing so, the linear polarization purity can reach $>$ 99\%, which can be derived by analyzing the incident and transmission lights of L$_m$ laser. The wavelength of the L$_m$ laser used in the experiment is measured with a precision of 100~MHz by a wavemeter (WS-7, HighFinesse GmbH). The power of L$_m$ laser is measured using a commercial power meter (S120VC, Thorlabs Inc.) with a power variation within 5\%. To increase the measurement accuracy, a "magic" magnetic field direction is chosen to make the magic wavelength insensitive to both the linear polarization purity and the polarization direction of the laser.

The ac Stark shift caused by a laser can be written in the form
\begin{equation}
\begin{split}
\Delta E_i=&-\frac{F^2}{2}\bigg[ \alpha_i^S(\omega)+A\cos\theta _k\frac{m_J}{2J}\alpha_i^V(\omega)\\
&+\frac{3\cos^2\theta_p-1}{2}\cdot\frac{3m_J^2-J(J+1)}{J(2J-1)}\alpha_i^T(\omega)\bigg], \label{eq1}
\end{split}
\end{equation}
where $F$ is the strength of the ac electromagnetic field, $\alpha_i^S(\omega)$, $\alpha_i^V(\omega)$, and $\alpha_i^T(\omega)$ are, respectively, the scalar, the vector, and the tensor polarizabilities for quantum state $i$ at frequency $\omega$, and the tensor component will be taken into account only when $J> 1/2$. Also in Eq.~(\ref{eq1}), the laser polarization $A$, the angle $\theta_k$ between the laser propagation direction $\hat{k}$ and the magnetic field direction $\hat{B}$, the angle $\theta_p$ between the laser polarization direction and $\hat{B}$ are all important parameters affecting the ac Stark shift. In previous theoretical calculations~\cite{Tang13,Kaur15,Jiang17}, $A=0$ and $\cos\theta_p = 1$ were chosen when calculating the polarizabilities and extracting the magic wavelengths under a linearly polarized laser field.

We first consider the case where $A=0$ and $\cos\theta_p = 1$ in our experiment. Unlike the 395~nm magic wavelength measurement, it is found that the magic wavelength here is very sensitive to the parameters $A$, $\theta_k$, and $\theta_p$. Thus, we have to make sure that these parameters are very stable and precise. The parameter $A$ is measured to be 0.005(5) that corresponds to an almost linear polarization, but the $A\cos\theta_k$ term still affects the measurement because the ac Stark shifts to the sublevels $m_J = -3/2$ and $m_J = 3/2$ are found to be different. Setting $\cos\theta_k$ to be 0 will lower the effect caused by the polarization impurity.

In the experimental setup, the L$_m$ laser polarization and propagation directions are kept unchanged. In the beginning of our measurement, the background magnetic field of the ion is compensated to 0  by adjusting the currents in the three pairs of Helmholtz coils. The magnetic field amplitude can be measured by observing the clock transition Zeeman components. By adjusting the currents in the coils, the relationship between the current in each pair of coils and the magnetic field it produces is measured. By changing the currents in the coils, one can produce the magnetic field of any direction while keeping the amplitude constant. In the end of our measurement, the compensated background magnetic field is measured again so that the background magnetic field drift amplitude can be evaluated.

To measure the magic wavelength $\lambda_m$, we studied ac Stark shift within a few nanometers around $\lambda_m$. We measured the ac Stark shifts at six wavelengths of L$_m$ laser, each being measured for about 2000~s. Then the six points were fitted linearly and the magic wavelength was obtained. Evaluation of systematic shifts is of great importance in the measurement of the infrared magic wavelength since it is sensitive to the above-mentioned parameters. The systematic shifts caused by the uncertainties in $\theta_k$ and $\theta_p$, by the laser power, by the broadband laser spectrum, and by the background magnetic field drift were also evaluated.

For estimating the systematic shift due to $\theta_p$, we scanned $\theta_p$ from $-30^{\circ}$ to $30^{\circ}$. We found that the measured magic wavelength became longer when $\theta_p$ was near 0, as observed in Ref.~\cite{Jiang17}.
Experimentally we can change $\theta_p$ until the measured magic wavelength becomes the longest. According to the precision of $\theta_p$ that we can experimentally have, $\theta_p$ could cause a measurement uncertainty of 0.03~nm. For estimating the systematic shift due to $\theta_k$ and $A$, we let the laser pass through a polarizer before it enters into the vacuum chamber. However, for some reasons, such as the viewports that would change the polarization slightly, we can still see strong effects caused by $A$. Technically, the magnetic field direction can be adjusted to make $\cos\theta_k = 0$. By scanning $\theta_k$, it was found that the measured magic wavelength was longer when $\theta_k$ was closer to 90$^{\circ}$. When measuring the magic wavelength difference between $m_J = 3/2$ and $m_J = -3/2$, we found that this difference came to 0 when
$\theta_k = 90^{\circ}$, indicating that the $A\cos\theta_k$ term no longer contributed to the systematic shift. Experimentally we can change $\theta_k$ until the measured magic wavelength difference between $m_J = 3/2$ and $m_J = -3/2$ becomes 0. The experimental precision of $\theta_k$ would cause a measurement uncertainty of 0.01~nm.

\begin{figure}
\includegraphics[width=8cm,angle=0]{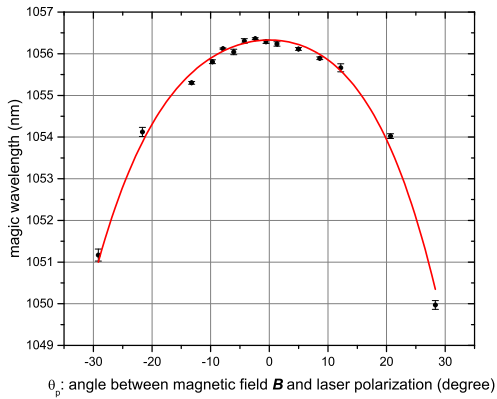}
\caption{The magic wavelength as a function of $\theta_p$. $\theta_p$ represents the angle between magnetic field direction and the laser polarization. Each data point shows the average of an experiment lasts for 1-4 hours. The error bars only include the statistical errors, yet the systematic errors caused by the magnetic field drifting, the laser power drifting, and the laser pointing drifting are not included. The fitted solid curve is a polynomial fit of the data set to the 4th order.} \label{fig1}
\vspace{0cm}
\end{figure}

The background magnetic field may be changing during the measurement. Since the measurement was found to be sensitive to the magnetic field direction, the effects of magnetic field change should be considered. By measuring the compensated magnetic field amplitude (which should be about 0) every few hours, the background magnetic field would only be changed by less than 30~nT during the whole experiment. Since the applied magnetic field amplitude is 3800~nT, we estimated that both $\theta_p$ and $\theta_k$ would gain an uncertainty of less than $0.5^{\circ}$ due to the background magnetic field change. According to the relationship between the magic wavelength and those parameters, magnetic field change during the whole experiment would cause a  magic wavelength measurement uncertainty of 0.08 nm.

Table~\ref{Tab1} lists the systematic error budget. Details about the systematic shift evaluation can be found in the Supplementary Materials.

\begin{table}
	\caption{Uncertainty budget for the infrared magic wavelength measurement. Effects with both shift and uncertainty smaller than 0.001~nm are not listed. Units are in nm.}
	\begin{center}
	\begin{tabular}{lccc}
		\hline\hline
		Source & Shift  &  Uncertainty   \\
		\hline
		Statistical         &   -        &  0.02              \\
		$\theta_p$          &   0        &  0.03              \\
		$\theta_k$          &   0        &  0.01              \\
		Laser power         &  $-0.03$     &  0.03              \\
		Broadband laser spectrum  &   0.005    &  0.005             \\
		Background magnetic field shift&   0        &  0.08              \\
		\hline
		Total uncertainty    &   $-0.04$    &  0.09              \\
		\hline
		Magic wavelength & & \\
		with correction & &1056.37(9)  \\
		\hline\hline
	\end{tabular}
    \end{center}
	\label{Tab1}
\end{table}

With the corrections shown in Table~\ref{Tab1}, the infrared magic wavelength for $|m_J|=3/2$ is determined as 1056.37(9)~nm. To date, there are a few theoretical calculations on this wavelength~\cite{Tang13,Kaur15,Jiang17}, as listed in Table~\ref{Tab2}. One can see that our result is in fairly good agreement with these calculations but with much smaller uncertainty.

Theoretically, using the perturbation theory, the dynamic electric dipole polarizabilities of a given atomic state can be expressed as
\begin{equation}
\begin{aligned}
\alpha_i^S&(\omega)=\frac{2}{3(2J_i+1)}\sum_k\frac{\Delta E_{ki}| \langle \Psi_i||D||\Psi_k \rangle|^2}{\Delta E_{ki}^2-\omega^2}\\
\alpha_i^V&(\omega)=\sqrt{\frac{24J_i}{(J_i+1)(2J_i+1)}}\\
&\times\sum_k(-1)^{(J_i+J_k+1)}\begin{Bmatrix}J_i&1&J_i \\ 1&J_k&1 \end{Bmatrix}\frac{\omega| \langle \Psi_i||D||\Psi_k \rangle |^2}{\Delta E_{ki}^2-\omega^2}\\
\alpha_i^T&(\omega)=\sqrt{\frac{40J_i(2J_i-1)}{3(J_i+1)(2J_i+1)(2J_i+3)}}\\
&\times\sum_k(-1)^{(J_i+J_k)}\begin{Bmatrix}J_i&2&J_i \\ 1&J_k&1 \end{Bmatrix}\frac{\Delta E_{ki}| \langle \Psi_i||D||\Psi_k \rangle |^2}{\Delta E_{ki}^2-\omega^2}
\label{eq2}
\end{aligned}
\end{equation}
where $D$ is the electric dipole transition operator.
It is noted that, when $\omega=0$, $\alpha_i^S(\omega)$, $\alpha_i^V(\omega)$, and $\alpha_i^T(\omega)$ are referred, respectively, as the static scalar, vector, and tensor polarizabilities for state $i$.  The uncertainties of the polarizabilities are governed by the uncertainties of the reduced transition matrix elements. Under our experimental conditions, the ac Stark shift at the magic wavelength includes the contributions from $\alpha_{4s}^S(\omega)$, $\alpha_{3d_{5/2}}^S(\omega)$, and $\alpha_{3d_{5/2}}^T(\omega)$, and the contribution from $\alpha^V(\omega)$ can be neglected.

Since the ac Stark shift of the clock transition at the magic wavelength is zero, the dynamic polarizabilities are the same for both $4s_{1/2}$ and $3d_{5/2}$ states. Theoretical works~\cite{Tang13,Safronova11} show that the contributions from the $4s_{1/2}\rightarrow 4p_{1/2}$ and $4s_{1/2}\rightarrow 4p_{3/2}$ transitions dominate the polarizability of the $4s_{1/2}$ state, and the contributions to the polarizability of the $3d_{5/2}$ state are dominated by the $3d_{5/2}\rightarrow 4p_{3/2}$ transition that constitutes over 80\% of the polarizability. Based upon the magic wavelength measured here, the energy levels of atomic states in Ca$^+ $ given by NIST~\cite{Kramida18}, the experimentally obtained high precision matrix elements for the $4s_{1/2}\rightarrow 4p_{1/2}$ and $4s_{1/2}\rightarrow 4p_{3/2}$ transitions~\cite{Liu15}, and other reduced matrix elements from RCC~\cite{Safronova11,Kaur17} and DFCP calculations, the matrix element $| \langle 3d_{5/2}||D||4p_{3/2} \rangle |$  is extracted to be 3.295(15)~a.u..

The BBR shift to the $4s_{1/2}\rightarrow 5d_{5/2}$ clock transition frequency can be evaluated according to
\begin{equation}
\begin{aligned}
\Delta_{\rm BBR}(4s_{1/2}\rightarrow5d_{5/2})=&-\frac{1}{2}(\alpha_{0,4s_{1/2}}-\alpha_{0,3d_{5/2}})\\
&\times(831.9{\rm V}/{\rm m})^2\bigg(\frac{T(\rm K)}{300}\bigg)^4  \label{eq3}
\end{aligned}
\end{equation}
where $\alpha_0$ is the static electric-dipole polarizability. Combining the matrix element $\left| \langle 3d_{5/2} \left\| D \right\|4p_{3/2} \rangle \right|$ obtained above and other matrix elements from both experiment and theoretical calculations, the differential static polarizability between the $4s_{1/2}$ and $3d_{5/2}$ states is determined to be $-44.32(32)$~a.u.. The corresponding BBR shift at 300~K is 0.3816(28)~Hz. Comparing to the existing theoretical values, as listed in Table~\ref{Tab2}, the present value agrees with and slightly better than the best previous theoretical calculation of  Ref.~\cite{Safronova11}. The fractional uncertainty of BBR shift can now be updated to be 6.8$\times 10^{-18}$. The uncertainty due to the knowledge of the dynamic polarizabilities can be further reduced with the method in Ref.~\cite{Barrett19}.

\begin{table}
	\caption{Comparison of the infrared magic wavelength (nm) and the Ca$^{+}$ blackbody radiation shift (Hz) at 300~K.}
	\begin{center}
	\begin{tabular}{lccccc}
		\hline\hline
		&Present & \multicolumn{2}{c}{Theory}\\
		\hline
        &   & All-order         &DFCP\\
& &method &method\\
        \hline
	Magic wavelength &1056.37(9) & 1052.26~\cite{Kaur15} & 1074(26)~\cite{Tang13} \\
& & &1074(32)~\cite{Jiang17}\\
			BBR shift  &0.3816(28)  & 0.3811(44)~\cite{Safronova11} & 0.380(14)~\cite{Arora07}\\
& & 0.31(1)~\cite{Sahoo09} & 0.368~\cite{Mitroy08}\\
			\hline\hline
		\end{tabular}
	\end{center}
	\label{Tab2}
\end{table}

In summary, we have performed an experimental determination of the infrared magic wavelength in Ca$^+$ with uncertainty less than $0.1$~nm. Our result agrees well with theoretical values but with 1-2 orders of magnitude improvement. By using our measured result, the differential static scalar polarizability has been determined as $-44.32(32)$~a.u., also in agreement with the previous theoretical values but with higher accuracy. The blackbody radiation shift at 300~K has then evaluated as 0.3816(28)~Hz, which is also in good agreement with our recent measurement~\cite{Huang19}. It is noted that the infrared magic wavelength for the $4s_{1/2}\rightarrow 5d_{5/2}$ transition ($m_J = 1/2$) was also predicted theoretically in Ref.~\cite{Tang13}. The matrix element of  $3d_{5/2}\rightarrow 4f_{7/2}$ transition, whose theoretical uncertainty is 1.1\% using relativistic all-order method, could be extracted and improved from further measurement on this magic wavelength, which can help reduce the BBR shift uncertainty further. Although the differential static scalar polarizability can be experimentally obtained with a better accuracy by measuring the magic rf field~\cite{Huang19} that could result in the BBR shift with lower uncertainty, it requires that the differential static polarizability of the clock transition is negative~\cite{Huang19,Dube14}. However, many optical clock candidates, such as Yb$^+$, In$^+$, Sr, and Yb, do not satisfy this criterion. The scheme in this work, which uses the magic wavelength to extract the transition matrix elements, can be an alternative and more general way to determine the differential static polarizability.

Furthermore, the determination of the infrared magic wavelength is also a very important step for building an all-optical trapping ion optical clock in the near future. Long-time all-optical trapping of the ions has already been achieved recently by Schatz's group~\cite{Lambrecht17}. It is found that one can trap an ion with optical dipole trap only if the trap potential is higher than the ion kinetic motion energy, and the heating rate of the dipole trap will be higher with relatively near resonance wavelength. The ion lifetime in dipole trap would be a few ms with a few hundreds of GHz red detuning lasers~\cite{Schneider10,Huber14}; yet the lifetime can be extended to a few second with a few hundred THz far-off-resonance lasers. To realize an ion-based optical clock with all-optical trapping scheme, lifetime of at least 100~ms is required and the heating rate should be maintained as low as possible in order to lower the Doppler and Stark shifts. Building a clock with infrared lasers of hundreds THz of red detuning is a better choice comparing to the 395~nm laser. Besides, one can easily obtain a fiber laser with higher power ($>$ 60 W) at the Ca$^+$ infrared magic wavelength in the range of $1000\sim1100$~nm. The all-optical trapping ion optical clock scheme can be used to trap multiple ions~\cite{Schmidt18}, which will potentially increase clock stability. However, The magic wavelength in sensitive to the alignment of the beam and its polarization relative to the magnetic field orientation, in our case, these effect would limit the precision of the magic wavelength to the 0.1 nm level, this would limit the accuracy of the optical clocks. In the practical point of view, building a high accuracy all optical ion clock would require techniques to make the laser pointing and magnetic field more stable.

We thank Jun Jiang, Yongbo Tang, Fangfei Wu, V. Yudin, A. Taichenachev, Zongchao Yan, B. Sahoo, and J. Ye for help and fruitful discussions. This work is supported by the National Key R\&D Program of China (Grant Nos. 2018YFA0307500, 2017YFA0304401, 2017YFA0304404, 2017YFF0212003), the Natural Science Foundation of China (Grant Nos. 11634013, 11622434, 91736310, 11774388), the Strategic Priority Research Program of the Chinese Academy of Sciences (Grant No. XDB21030100), CAS Youth Innovation Promotion Association (Grant Nos. 2015274, 2018364), and Hubei Province Science Fund for Distinguished Young Scholars (Grant No. 2017CFA040).



\end{document}